\documentclass[twocolumn,prl,10pt,aps,floatfix]{revtex4-1}
\usepackage{graphicx}
\usepackage{amsmath}
\usepackage[colorlinks,linkcolor=blue,citecolor=blue,urlcolor=blue]{hyperref}
\usepackage{color}

\begin{document}
\title{Spin Seebeck effect in the classical easy-axis 
antiferromagnetic chain}  
\author{X. Zotos$^{1,2}$}
\affiliation{$^1$Department of Physics,
University of Crete, 70013 Heraklion, Greece}
\affiliation{$^2$
Max-Planck-Institut f\"ur Physik Komplexer Systeme, 01187 Dresden, Germany} 

\date{\today}

\begin{abstract}
By molecular dynamics simulations we study  the spin Seebeck effect 
as a function of magnetic field 
in the prototype classical easy-axis antiferromagnetic chain, in the far-out 
of equilibrium as well as linear response regime. 
We find distinct behavior in the low field antiferromagnetic, middle field 
canted and high field ferromagnetic phase. In particular,
in the open boundary system at low temperatures, we observe
a divergence of the spin current in the spin-flop transition between the 
antiferromagnetic and canted phase, accompanied by a 
change of sign in the generated spin current by the 
temperature gradient. These results are corroborated by a simple spin-wave 
phenomenological analysis and simulations in the linear response regime.
They shed light on the spin current sign change observed in experiments in 
bulk antiferromagnetic materials.
\end{abstract}
\maketitle

\section{Introduction}
The generation and control of spin currents is a central topic in the field 
of spintronics \cite{review}. 
In particular the spin Seebeck effect \cite{kikkawa}, 
the generation of a spin current by a temperature gradient in a magnetic field, 
has been extensively experimentally and 
theoretically studied in  a great variety of bulk magnetic 
systems as for instance, 
the ferrimagnetic YIG/Pt heterostructures, antiferromagnetic materials
(e.g. Cr$_2$O$_3$, Fe$_2$O$_3$) and Van der Vaals 2D materials as the 
quasi-2D layered ferromagnets, Cr$_2$Si$_2$Te$_6$ and Cr$_2$Ge$_2$Te$_6$ 
(for an extensive reference \cite{review}).
In particular, concerning easy-axis bulk antiferromagnetic materials,  
there is  experimental \cite{wu,li,li2,yuan} and theoretical 
\cite{rezende, yamamoto, reitz, yamamoto2} 
interest and debate concerning the sign of the generated 
spin current \cite{kikkawa, tserkovnyak}. 

In a different research domain, the physics of (quasi-) one dimensional 
magnetic systems, both classical and quantum, has been studied for years, 
starting with the Bethe ansatz solution of the antiferromagnetic 
spin-1/2 chain.
In particular, the  exotic physics of easy-axis antiferromagnetic spin chains 
\cite{mikeska}
and quantum spin liquid materials with topological 
spinon excitations has attracted great interest.
The spin Seebeck effect in the spin-1/2 chains Sr$_2$CuO$_3$ with spinon and 
CuGeO$_3$ with triplon excitations  
has been studied experimentally \cite{hirobe, triplon} and rigorously 
evaluated theoretically \cite{psaroudaki} in the easy-plane regime.
Furthermore, the thermal 
transport of classical spin chains has been studied 
by  numerical dynamics simulations \cite{savin} and
lately, the character of spin transport, ballistic, diffusive or anomalous, 
in classical and quantum (anti-) ferromagnetic chains attracts a great 
deal of attention (for a recent tour de force and references therein see 
\cite{google}).  

The spin Seebeck effect has never been studied for 
the easy-axis classical antiferromagnetic spin chain and it 
makes sense to try to understand the physics of the effect in this 
prototype but realistic model.
It allows us to clarify the relation of the 
sign of the induced spin current by a temperature gradient 
across the spin-flop transition occuring at a critical field and
serves as a bridge between spintronics studies in bulk materials 
and model magnetic systems.
Besides the academic interest, quasi-one dimensional compounds exist 
that can offer a platform for obtaining spin currents 
besides the bulk materials usually studied.

In the following, we first employ standard 
molecular dynamics (MD) simulations \cite{md} to study 
the out of equilibrium spin current generation by a thermal current in a 
magnetic field. We find a sign reversal of the spin current 
at the critical field between the antiferromagnetic and 
canted ferromagnetic phase that we analyze by a simple spin-wave theory.
The divergence of the  induced current 
at the spin-flop transition, 
could be observed in large spin quasi-one dimensional 
spin chain compounds.
The picture of the far-out of equilibrium spin Seebeck effect is 
corroborated by a simple spin-wave phenomenological model and 
simulations  in the linear response regime.

\section{Model}
The model we study is the classical antiferromagnetic Heisenberg chain
with easy-axis anisotropy given by the Hamiltonian,
\begin{equation}
H=\sum_{l=1}^L J_{\perp} (
S^x_l S^x_{l+1} + S^y_l S^y_{l+1}) +
\Delta S^z_l S^z_{l+1} -h S^z_l,
\label{model}
\end{equation}
\noindent
where ${\bf S_l}$ is a unit vector with components $S^{x,y,z}_l$,
$J_{\perp}>0$ is the in-plane and $\Delta >0$ the easy-axis 
exchange interactions with $\Delta > J_{\perp}$ and $h$ the magnetic field.
We will use the common parametrization 
$S^x_l=\sin \theta_l\cdot \cos \phi_l, S^y_l=\sin \theta_l\cdot \sin \phi_l,
S^z_l=\cos \theta_l$.

\noindent
The spin dynamics is given by 
Landau-Lifshitz equations of motion,
\begin{equation}
\frac{d}{dt}{\bf S_l}={\bf S_l}\times
\Big(- \frac{\partial H}{\partial{\bf S_l}}\Big).
\label{ll}
\end{equation}

To study far-out of equilibrium transport, 
we use a straightforward numerical method, 
simulating the microscopic heat transfer by 
embedding the spin system between two Langevin thermostats at 
temperatures $T_L, T_R$, realized by 
two Heisenberg chains of length $N_L, N_R$.
We apply the Heun method \cite{md} 
to numerically integrate the stochastic version 
of the Landau-Lifshitz-Gilbert equation for magnetic systems,

\begin{equation}
(1+\alpha^2)\frac{d}{dt}{\bf S_l}=
{\bf S_l}\times
(\xi_l- \frac{\partial H}{\partial{\bf S_l}})-
\alpha {\bf S_l}\Big[ {\bf S_l}\times 
(\xi_l- \frac{\partial H}{\partial{\bf S_l}})\Big]
\label{md}
\end{equation}
\noindent
where $\alpha$ is a damping coefficient and $\xi_l$ a white Gaussian noise
representing the thermostat at temperature $T$,
\begin{equation*}
\langle\xi_l(t)\rangle=0,~~~\langle\xi_l(t_1)\xi_k(t_2)\rangle
=2\alpha T\delta_{lk}\delta(t_1-t_2).
\end{equation*}

The spin $J^S$ and energy $J^E$ currents are  given by the 
corresponding spin and energy continuity equations \cite{nz,meisner},
\begin{equation}
J^S=\sum_l J_{\perp} (S^x_l S^y_{l+1}-S^y_l S^x_{l+1})
\label{js}
\end{equation}

\begin{eqnarray}
J^E=-\sum_l
&&J_{\perp}^2 (S^x_{l-1}S^z_l S^y_{l+1}-S^y_{l-1}S^z_l S^x_{l+1})
\nonumber\\
&-&J_{\perp}\Delta S^z_{l-1}(S^x_l S^y_{l+1}-S^y_l S^x_{l+1})
\nonumber\\
&-&J_{\perp}\Delta (S^x_{l-1} S^y_l-S^y_{l-1} S^x_{l})S^z_{l+1}.
\label{je}
\end{eqnarray}

We first establish the phase diagram, in the zero temperature limit, 
by considering the high field 
region where $\theta_l=\theta,~~\phi_{l+1}-\phi_l=\pi$, obtaining
by minimization of the energy, 

\begin{equation*}
E_{ferro}=-J_{\perp} \sin^2 \theta+\Delta \cos^2\theta -h \cos \theta
\end{equation*} 

\begin{equation}
z\equiv\cos \theta=\frac{h}{2(J_{\perp}+\Delta)}.
\end{equation}

\noindent
The critical field $h_f=2(J_{\perp}+\Delta)$ above which we 
have the ferromagnetic phase is obtained setting  $z=1$.
The critical magnetic field $h_c$, above which we have a 
canted ferromagnetic phase and below an antiferromagnetic one
with $\theta_{l+1}-\theta_l=\pi$, 
is obtained by equating the energies of the two states
 
\begin{equation*}
E_{\rm ferro}=E_{\rm afo}=-\Delta,~~~ h_c=2\sqrt{\Delta^2-J_{\perp}^2}.
\end{equation*}

\section{MD results}
\begin{figure}[!h]
\begin{center}
\includegraphics[width=1\linewidth, angle=0]{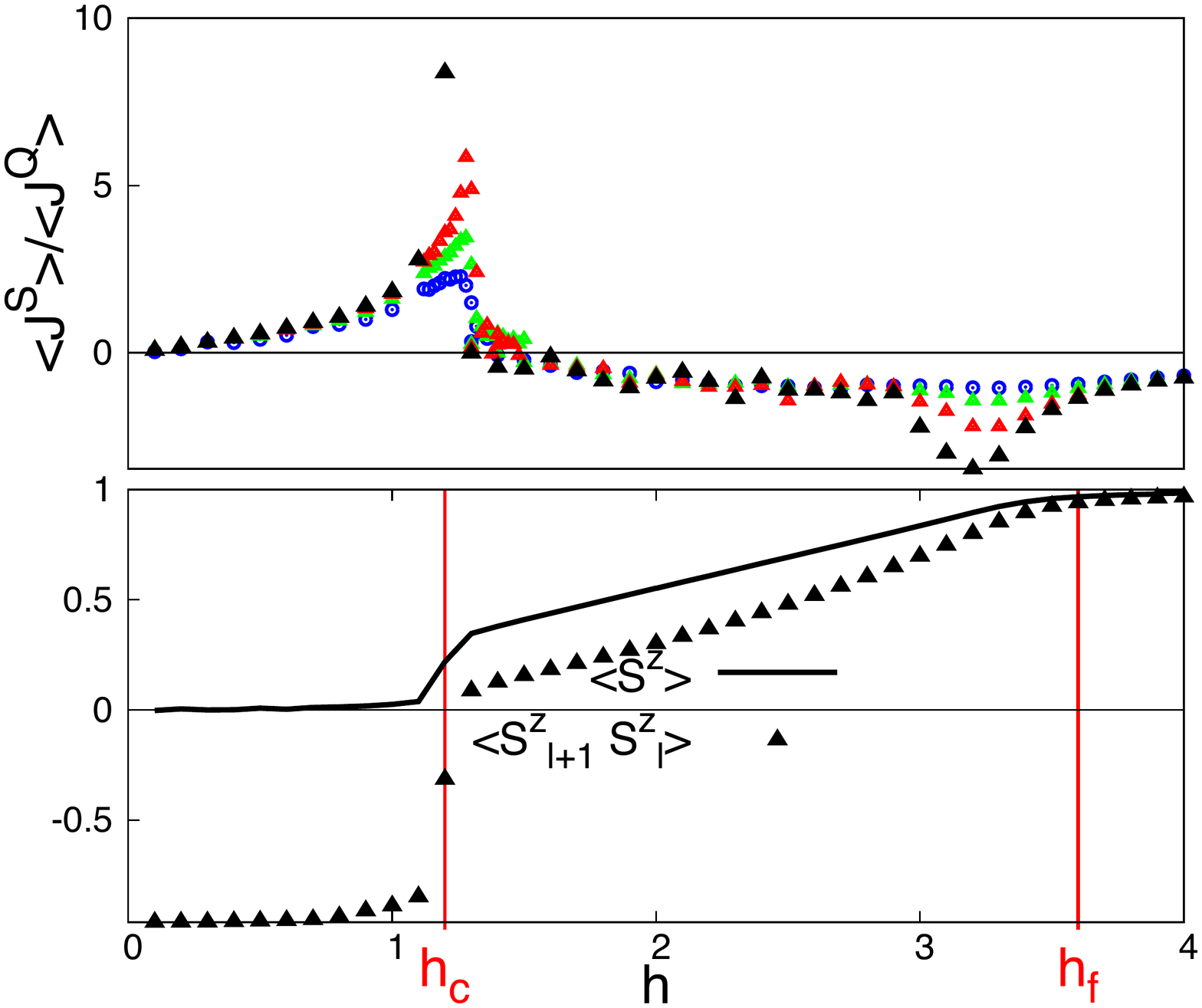}
\caption{Ratio of spin to thermal current as a function of magnetic 
field for $J_{\perp}=0.8$. The mean $(T_L+T_R)/2$ temperature is $T=0.02$ 
and the system sizes, 
$L=160~({\rm blue}), 320~({\rm green}), 640~({\rm red}), 1280~({\rm black})$. 
Also shown below the mean magnetization (black line) and nearest neighbor 
spin-spin correlation (black triangles).}
\label{fig1}
\end{center}
\end{figure}

In Fig.\ref{fig1} we show the ratio of the mean spin current 
$\langle J^S\rangle$ to 
the mean thermal current 
$\langle J^Q\rangle=\langle J^E\rangle-h \langle J^S\rangle$.
The mean values, $\langle O\rangle =\frac{1}{L}\langle \sum_l O_l\rangle$, 
$O_l$ a local quantity, 
are obtained by averaging over about $10^8$ samples by sweeping 
over all lattice sites (we take, $N_L=N_R=L/2$). 
The thermal current is
induced by setting the left (right) baths at different temperatures
$T_L (T_R)$, creating a constant temperature gradient along the chain.  
In the middle of the chain ($l=1,L$)  
the damping coefficient $\alpha$ and the white Gaussian noise
$\xi_l$ are set equal to zero.
The mean temperature is $T=0.02$, with up to $T_L=0.03,~~T_R=0.01$. 
Here, $J_{\perp}=0.8$ and we take 
$\Delta=1$ as the unit of energies and temperature,  implying
critical fields $h_c=1.2,~h_f=3.6$.

Concerning the numerical simulation, 
we find that the results are essentially independent of the temperature 
gradient within the accuracy of the simulation.
The thermal gradient induces a thermal current, which in turn induces 
a spin current. Being a second order effect, the measured spin current  
shows rather large fluctuations in the data compared to the thermal current.
Thus we use relatively large 
temperature gradients to improve the accuracy of the data. 
In the particular simulations we used as baths isotropic 
antiferromagnetic Heisenberg chains ($J_{\perp}=\Delta=1$)
in zero magnetic field for which the energy-temperature relation 
is known. However, 
we found that the use of other baths, e.g. a ferromagnetic chain, or a phonon 
bath, does not qualitatively change the spin current-thermal current relation. 

The most notable feature in the data in Fig.\ref{fig1} 
is the sharp reversal of the spin current 
at the spin-flop transition (to be dicussed  
later in the framework of a spin wave theory)
and the size dependence indicating a diverging spin current in the zero 
temperature limit. 
In the low field antiferromagnetic phase the spin current is 
in the same direction as the thermal current, while in the ferromagnetic one 
it is opposite to the thermal current. Of course, as expected, reversing the 
direction of the magnetic field, reverses this relation.
In the second part of the figure (below), we show the magnetic field 
dependence of the 
average magnetization $\langle S^z\rangle$ 
and nearest neighbor spin correlation 
$\langle S^z_{l+1}S^z_l\rangle$,
clearly indicating the development of the three magnetic phases. 
 
\begin{figure}[!h]
\begin{center}
\includegraphics[width=1\linewidth, angle=0]{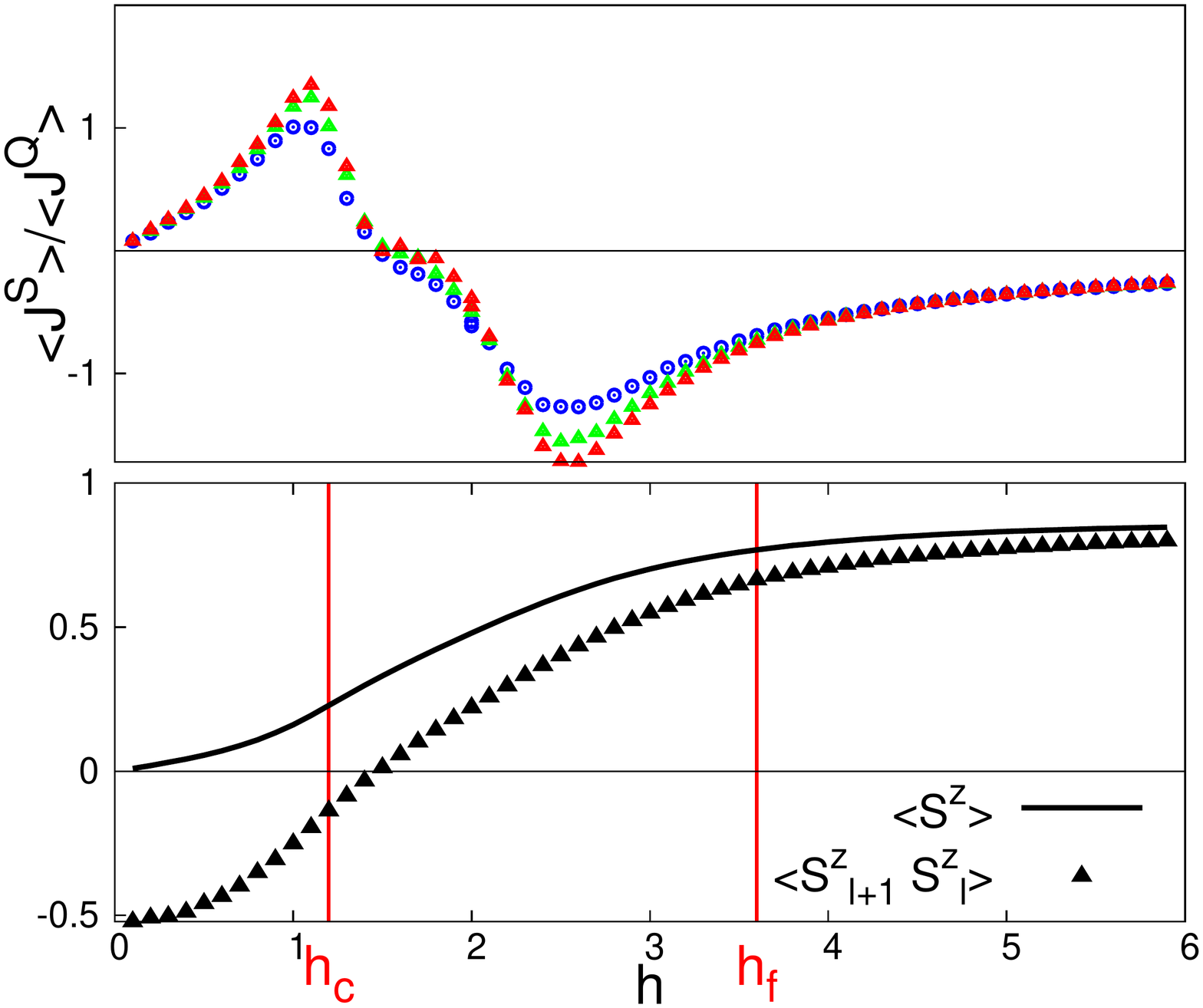}
\caption{Ratio of spin to thermal current as a function of magnetic 
field for $J_{\perp}=0.8$. The mean $(T_L+T_R)/2$ temperature is $T=0.2$ 
and the system sizes, 
$L=160~({\rm blue}), 320~({\rm green}), 640~({\rm red})$. 
Also shown below the mean magnetization (black line) and nearest neighbor 
spin-spin correlation (black triangles).}
\label{fig2}
\end{center}
\end{figure}

In Fig.\ref{fig2}, we show 
the same quantities at a higher temperature 
where the transitions are smoothed out but the same features remain.
Also, the finite size effects are reduced as well as 
the ratio of the spin to thermal
current, in relation to the temperature increase. 

\begin{figure}[!h]
\begin{center}
\includegraphics[width=1\linewidth, angle=0]{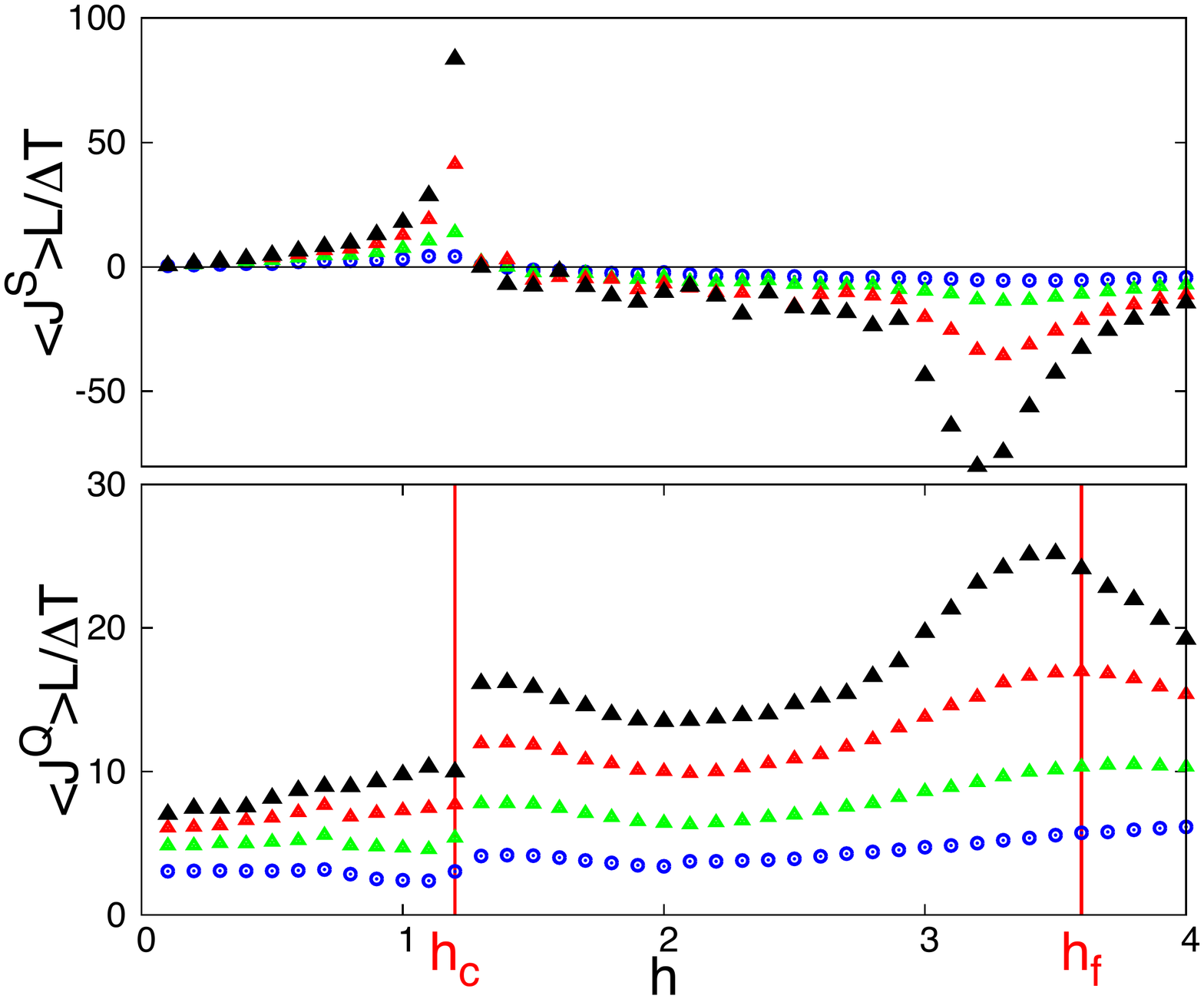}
\caption{Spin and thermal conductivities as a function of magnetic 
field for $J_{\perp}=0.8$. The mean $(T_L+T_R)/2$ temperature is $T=0.02$ 
and the system sizes, 
$L=160~({\rm blue}), 320~({\rm green}), 640~({\rm red}), 1280~({\rm black})$.} 
\label{fig3}
\end{center}
\end{figure}

In Fig.\ref{fig3} we show the field dependence of the 
spin and thermal conductivities separately. 
While the thermal current shows anomalies at 
the spin-flop  and ferromagnetic transition, 
the spin current is clearly responsible for the sign changes and overall 
behavior shown in Fig.\ref{fig1}.

\begin{figure}[!h]
\begin{center}
\includegraphics[width=1\linewidth, angle=0]{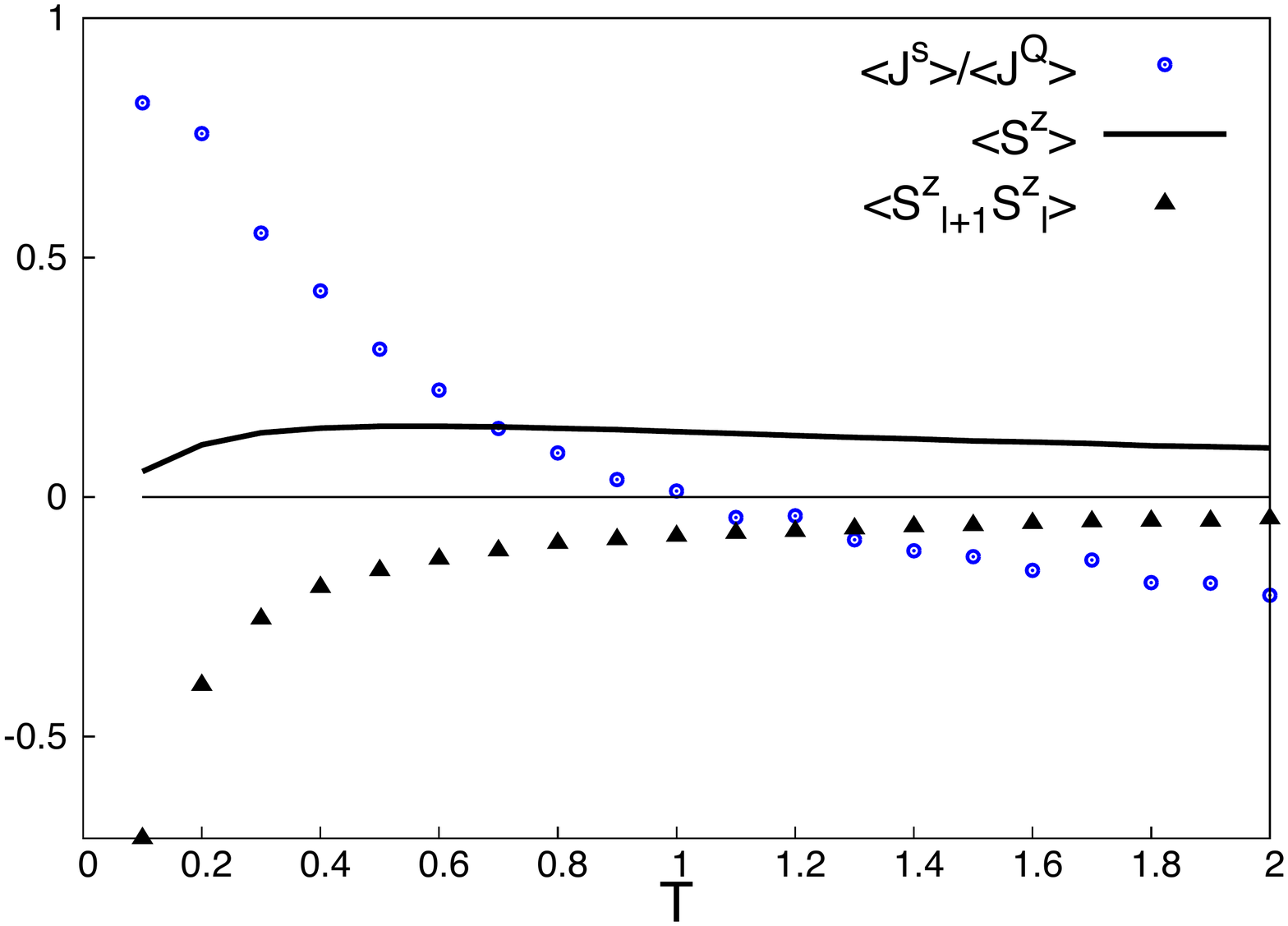}
\caption{Ratio of spin to thermal current, magnetization and nearest neighbor 
correlations as a function of temperature 
for a magnetic field $h=0.8$ and $L=160$.}
\label{fig4}
\end{center}
\end{figure}

Finally in Fig.\ref{fig4} we show the temperature dependence of the 
spin Seebeck effect  by the ratio $\langle J^S\rangle/\langle J^Q\rangle$ 
at $h=0.8$. 
In this field we are at  the 
antiferromagnetic regime at low temperatures and the sign of the ratio is 
positive. Raising the temperature, the antiferromagnetic phase "melts" 
with the appearance of an increasing number of domain walls, till  
a critical temperature $T\sim 1$ where we observe a change to 
a negative sign of the spin current,
as in the ferromagnetic regime. To get an insight to this picture 
we show the temperature dependence of the uniform magnetization, 
rather small at this field 
and the decreasing nearest neighbor antiferromagnetic spin-spin correlations.

\section{Spin-wave analysis}
We can reach an understanding of the spin current sign reversal 
and divergence at the spin-flop  transition,
by considering a simple linear spin-wave theory.
First, in the high field $|h| >h_c$ canted ferromagnetic phase, linearizing 
in $S^x_l, S^y_l$ the equations of motion (\ref{ll}), we obtain,
\begin{eqnarray*}
{\dot S^x}_l&=& -S^y_l(2\Delta z-h)+z J_{\perp} (S^y_{l+1}+ S^y_{l-1})
\nonumber\\
{\dot S^y}_l&=& +S^x_l(2\Delta z-h) -zJ_{\perp}(S^x_{l+1}+S^x_{l-1})],
\end{eqnarray*}
\noindent
(the dot indicating the time derivative).
With the substitution,

\begin{equation}
S^x_l \pm i S^y_l=ue^{i(ql-\omega_{\pm} t)},
\label{ferrosxsy}
\end{equation}

\noindent
we obtain the spin-wave spectrum,

\begin{equation*}
\omega_{\pm} =\pm h \mp 2z(\Delta-J_{\perp} \cos q)
\end{equation*}

\noindent
with positive eigenfrequencies,

\begin{equation}
\omega=h\Big(1-\frac{\Delta-J_{\perp}\cos q}{J_{\perp}+\Delta}\Big).
\label{wferro}
\end{equation}

\noindent
Using (\ref{js},\ref{je},\ref{ferrosxsy}), by substituting 
the values of $S^{x,y,z}_l$ for a spin-wave of wavevector $q$ 
we obtain for the spin and thermal current per unit length, 
\begin{eqnarray*}
j^s_q&=&J_{\perp} u^2\sin q,
\nonumber\\
j^{\epsilon}_q&=&=
-2J_{\perp} u^2(J_{\perp}\cos q-\Delta )z\cdot \sin q
\nonumber\\
u^2&=&1-z^2
\nonumber\\
j^Q_q&=&j^{\epsilon}_q-h j^s_q
\nonumber\\
j^Q_q&=&-
\Big(\frac{J_{\perp}(1+\cos q)}{\Delta+J_{\perp}}\Big)\cdot h j^s_q.
\end{eqnarray*}

\noindent
Thus we see that in the high field  region, for $h>0$ the
spin and thermal current 
have opposite sign, $j^Q_q > 0$,~~$j^s_q <0$
and of course for $h<0,~~ j^Q_q > 0,~~j^s_q >0$.

In the low field antiferromagnetic region, the equations of motion are,
\begin{eqnarray*}
{\dot S^x}_{2l}&=& -S^y_{2l}(2\Delta S^z_{odd} -h)
+S^z_{even}J_{\perp}(S^y_{2l+1}+S^y_{2l-1})
\nonumber\\
{\dot S^y}_{2l}&=&+S^x_{2l}(2\Delta S^z_{odd} -h)
-S^z_{even}J_{\perp}(S^x_{2l+1}+S^x_{2l-1})
\nonumber\\
{\dot S^x}_{2l+1}&=&-S^y_{2l+1}(2\Delta S^z_{even}-h)
+S^z_{odd}J_{\perp}(S^y_{2l}+S^y_{2l+2})
\nonumber\\
{\dot S^y}_{2l+1}&=&+S^x_{2l+1}(2\Delta S^z_{even}-h)
-S^z_{odd}J_{\perp}(S^x_{2l}+S^x_{2l+2}),
\end{eqnarray*}

\noindent
with $S^z_{odd,even}$ the alternating $S^z$ component at the odd, even sites.
With the substitution,
\begin{eqnarray*}
S^x_{2l}\pm i S^y_{2l}&=&u_{\pm} e^{i q 2l-\omega_{\pm}t}
\nonumber\\
S^x_{2l+1}\pm i S^y_{2l+1}&=&v_{\pm} e^{i q (2l+1)-\omega_{\pm}t}
\end{eqnarray*}

\noindent
and taking e.g. $S^z_{even}=+1, S^z_{odd}=-1$, 
we obtain the eigenvalue problem, 

\begin{equation*}
\begin{pmatrix}
\pm (2\Delta +h)& \pm 2J_{\perp}\cos q \\
\mp 2J_{\perp}\cos q & \mp (2\Delta-h )\\
\end{pmatrix}
\begin{pmatrix}
u_{\pm}\\
v_{\pm}
\end{pmatrix}
=\omega_{\pm}
\begin{pmatrix}
u_{\pm}\\
v_{\pm}
\end{pmatrix}
\end{equation*}

\noindent
for the frequency spectrum,

\begin{eqnarray}
\omega_{+\pm}&=&+ h\pm 2\sqrt{\Delta^2-J_{\perp}^2 \cos^2 q}
\nonumber\\
\omega_{-\pm}&=&- h\pm 2\sqrt{\Delta^2-J_{\perp}^2 \cos^2 q}.
\label{wafo}
\end{eqnarray}

\noindent
The positive frequency dispersions are,
$\omega_{\pm +}=\pm h+ h_c\sqrt{\frac{\Delta^2-J_{\perp}^2 \cos^2 q}
{\Delta^2-J_{\perp}^2}}$. The lower frequency dispersion 
$\omega_{-+}$ vanishes as $q\rightarrow 0$ at the 
critical field $h_c$, signaling the spin-flop transition.

Setting $u_--=\cos \phi,~v_-=\sin \phi,~
\tan \phi=\frac{-\Delta-\sqrt{\Delta^2-J_{\perp}^2\cos^2 q}}
{J_{\perp}\cos q}$, we obtain the currents for the lower frequency dispersion,
\begin{eqnarray*}
j^s_q&=&-J_{\perp} u_-\cdot v_-\sin q=J_{\perp} \frac{J_{\perp}\cos q}
{2\Delta} \sin q
\nonumber\\
j^{\epsilon}_q&=&\frac{1}{2}J_{\perp}^2 (v_-^2-u_-^2) \sin 2q
\nonumber\\
&=&\frac{1}{2}J_{\perp}^2 \frac{\sqrt{\Delta^2-J_{\perp}^2\cos^2 q}}{\Delta}
\sin 2q
\nonumber\\
\end{eqnarray*}
\begin{eqnarray}
j^Q_q&=&j^s\cdot (h_c\cdot \sqrt{1+\frac{J_{\perp}^2(1-\cos^2 q)}
{\Delta^2-J_{\perp}^2}}-h),
\nonumber\\
j^s_q/j^Q_q &>&0.
\label{ajsjq}
\end{eqnarray}

For the higher frequency dispersion $\omega_{++}$, setting,
$u_+=\cos \phi,~v_+=\sin \phi,~
\tan \phi=\frac{-\Delta+\sqrt{\Delta^2-J_{\perp}^2\cos^2 q}}
{J_{\perp}\cos q}$, we obtain the currents,
\begin{eqnarray*}
j^s_q&=&J_{\perp}u_+v_+\sin q=-J_{\perp} \frac{J_{\perp}\cos q}
{2\Delta} \sin q
\nonumber\\
j^{\epsilon}_q&=&
\frac{1}{2}J_{\perp}^2 (u_+^2-v_+^2) \sin 2q
\nonumber\\
&=&\frac{1}{2}J_{\perp}^2 \frac{\sqrt{\Delta^2-J_{\perp}^2\cos^2 q}}{\Delta}
\sin 2q
\nonumber\\
\end{eqnarray*}
\begin{eqnarray*}
j^Q_q&=&j^s\cdot (-2\sqrt{\Delta^2-J_{\perp}^2\cos^2 q}-h)
\nonumber\\
j^s_q/j^Q_q &<&0.
\end{eqnarray*}

Thus, in the antiferromagnetic region, as observed in the simulations above,
for $|h| < h_c$ the spin and thermal current of the dominating lower frequency 
dispersion spin-waves have the same sign $j^s_q/j^Q_q >0$. 
We can also get a hint on the diverging behavior of 
$\langle J^s\rangle/\langle J^Q\rangle$ 
for $h\rightarrow h_c$ from (\ref{ajsjq}) as at low energies 
for $q\rightarrow 0$ this ratio diverges. 

\section{''Landauer" approach}
The classical Heisenberg chain is a strongly interacting model with 
nonlinear equations of motion describing the spin dynamics.
Therefore we expect normal transport coefficients \cite{savin}
e.g. finite 
thermal and spin conductivity, due to spin wave - spin wave scattering,
although the anomalous behavior of spin transport in the isotropic Heisenberg 
model is presently in the focus of many theoretical studies \cite{google}.
 
\begin{figure}[!h]
\begin{center}
\includegraphics[width=1\linewidth, angle=0]{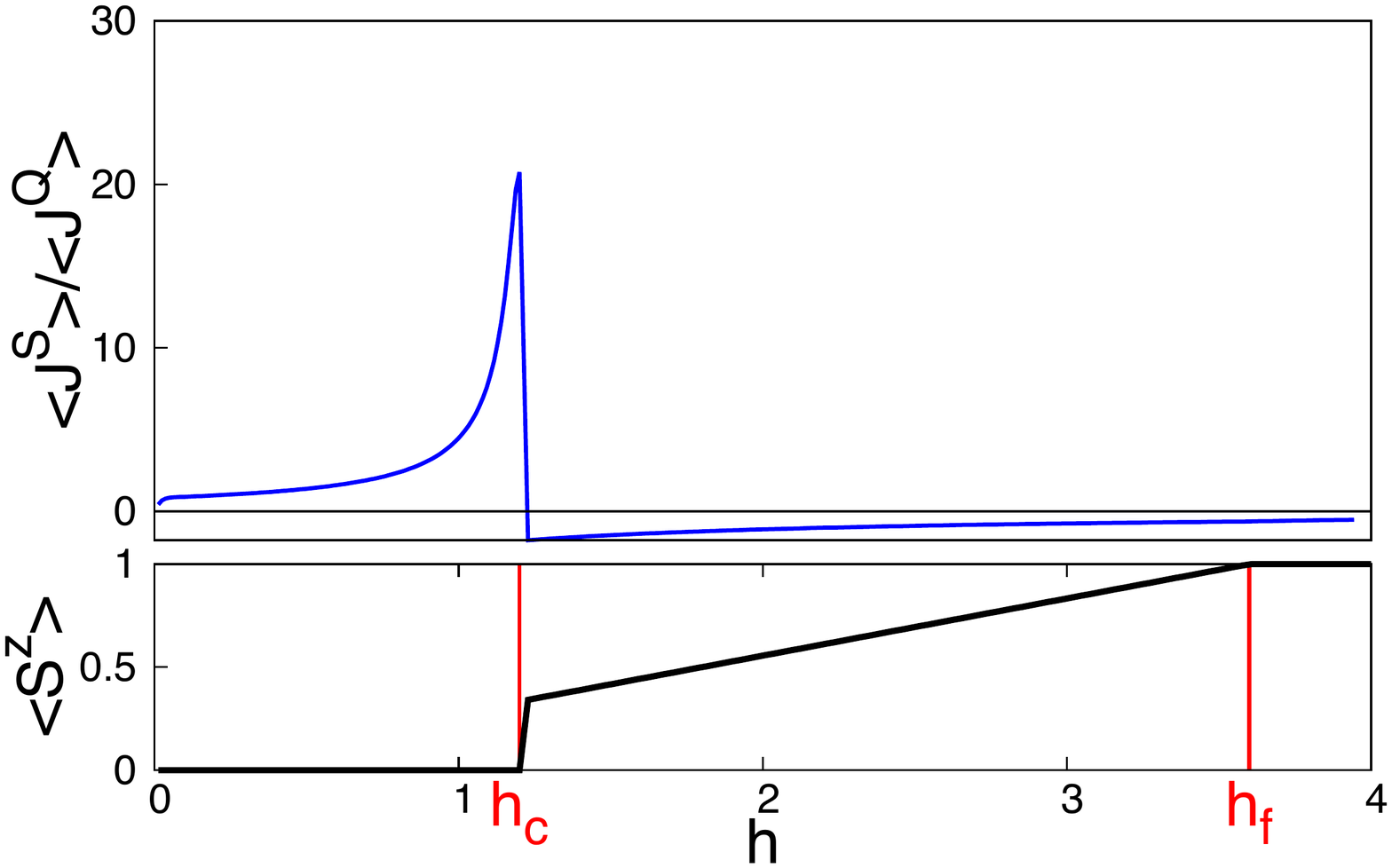}
\caption{Spin current to thermal current ratio and magnetization as 
a function of magnetic field for $J_{\perp}=0.8$. 
Average temperature $T=(T_L+T_R)/2=0.02$.}
\label{fig5}
\end{center}
\end{figure}

Nevertheless, for this open system with baths, 
we can obtain a heuristic description, shown in  Fig.\ref{fig5}, 
of the spin to thermal current ratio over the whole phase diagram 
by considering a phenomenological "Landauer" type model.
This can be justified by the low temperature in the simulations which 
implies a low spin wave density.

Assuming that spin and energy currents are emitted at the left-right leads 
at temperatures $T_{L,R}, (\beta_{L,R}=1/T_{L,R})$, we obtain for $h<h_c$ 
(antiferromagnetic region), 
\begin{eqnarray*}
\langle J^S\rangle&=&\sum_{\pm}\int_0^{+\pi/2}\frac{dq}{2\pi} 
(n_{\pm q}^L-n_{\pm q}^R)j^s_{\pm q}
\nonumber\\
\langle J^Q\rangle&=&\sum_{\pm}\int_0^{+\pi/2}\frac{dq}{2\pi} 
(n_{\pm q}^L-n_{\pm q}^R)j^Q_{\pm q}
\end{eqnarray*}

\noindent
summing the contributions over the positive frequency dispersions (\ref{wafo}) 
with $n_{\pm q}^{L,R}=
1/(e^{\beta_{L,R}\omega_{\pm q}}-1)$ (here we assume for simplicity a 
bosonic thermal distribution function).
Similar expressions are obtained for $h>h_c$ summing over the positive 
frequency dispersion (\ref{wferro}).

\section{Linear response}
Last but not least, in linear response the spin and thermal currents are
related by transport coefficients $C_{ij}$, 
\begin{equation}
\begin{pmatrix} J^Q \\ J^S \end{pmatrix} =
\begin{pmatrix} C_{QQ} & C_{QS} \\ C_{SQ} & C_{SS} \end{pmatrix}
\begin{pmatrix} -\nabla T \\ \nabla h \end{pmatrix}\,,
\label{MatrixEquation}
\end{equation}
where $C_{QQ}=\kappa_{QQ}$ ($C_{ss}=\sigma_{ss}$) is the heat (spin) 
conductivity. The coefficients $C_{ij}$ are given by the thermal average of 
time-dependent current-current correlation functions in a closed 
system with periodic boundary conditions,
\begin{equation*}
C_{i,j}=\frac{1}{L}\int_0^{\infty} dt \langle J^i(t)\cdot J^j(t=0)\rangle.
\end{equation*}

\begin{figure}[!h]
\begin{center}
\includegraphics[width=1\linewidth, angle=0]{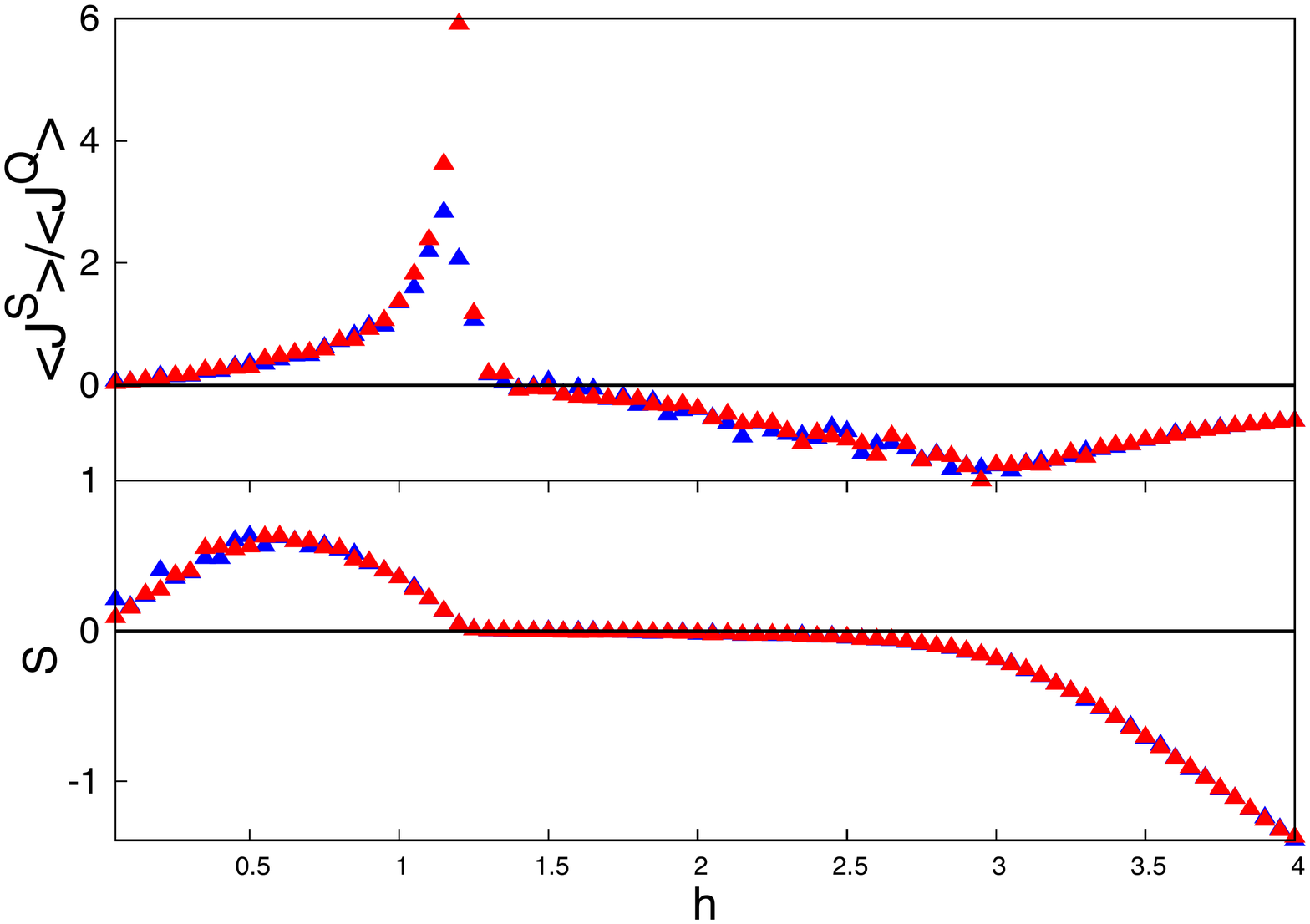}
\caption{Linear response coefficients as 
a function of magnetic field at $J_{\perp}=0.8$ for $L=160$ (blue) and 
$L=320$ (red) at temperature $T=0.05$.}
\label{fig6}
\end{center}
\end{figure}

\noindent
The time dependence is obtained by the same molecular dynamics procedure 
(\ref{md}) after equilibrating the system at a given temperature and 
then switching-off the thermal noise.
In Fig.\ref{fig6} we show two situations, (i) a system with no 
spin accumulation by setting $\nabla h=0$, relevant to an open system 
and (ii) a system 
with no spin current, $\langle J^S\rangle=0$ 
giving the spin Seebeck coefficient 
$S=\frac{\nabla h}{\nabla T}=\frac{C_{SQ}}{C_{SS}}$.
For the open system, we find the same behavior of $\langle J^S\rangle/
\langle J^Q \rangle$ as in the MD simulations.

\section{Conclusions}
We have studied the spin Seebeck effect in the most 
simple prototype classical easy-axis magnetic chain model by molecular dynamics 
simulations and basic spin wave theory.
We have found a sign change at the spin flop transition and 
clarified the role of spin wave excitations
in the low field antiferromagnetic phase as well as 
in the high field ferromagnetic phase.
This classical model could be realized experimentally in quasi-one 
dimensional large spin compounds but also provides a guide to the 
spin Seebeck effect studied over many years in bulk magnetic materials. 
The observations of this study should be extended to quantum spin systems, 
as the spin-1/2 easy-axis Heisenberg model, where the integrability of the 
model \cite{psaroudaki} allows an exact evaluation 
of the spin Seebeck coefficient. 
The scope is to assess the potential of the large variety of 
spin chain materials for spintronic applications.
 
\section{Acknowledgments}
X.Z. acknowledges stimulating discussions with Profs.  C. Hess, 
P. van Loosdrecht and M. Valldor.

\end{document}